\documentstyle[preprint,aps]{revtex}
\tightenlines

\newfont{\lfig}{cmr10 scaled\magstep0} 
\begin{document}
\draft
\title{Chiral Dynamics and the $S_{11}(1535)$ Nucleon Resonance}
\author{N. Kaiser, P.B. Siegel$^{**}$, and W. Weise}
\address{Physik Department, Technische Universit\"{a}t M\"{u}nchen\\
   Institut f\"{u}r Theoretische Physik, D-85747 Garching, Germany}

\bigskip

\maketitle

\begin{abstract}
The $SU(3)$ chiral effective lagrangian at next-to-leading order is applied to
the S-wave meson-baryon interaction in the energy range around the $\eta N$
threshold. Potentials are derived from this lagrangian and used in
a coupled channel calculation of the $\pi N$, $\eta N$, $K \Lambda$, $K \Sigma$
system in the isospin-$1/2$, $l=0$ partial wave.  Using the same parameters as
obtained from a fit to the low energy $\overline{K} N$ data it is found that a
quasi-bound $K \Sigma$-state is formed, with properties remarkably similar to
the  $S_{11}(1535)$ nucleon resonance. In particular, we find a large partial
decay width into $\eta N$ consistent with the empirical data.
\end{abstract}

\vspace{3in}

\noindent $^*${\it Work supported in part by BMBF and GSI}

\noindent $^{**}${\it On sabbatical leave from California State Polytechnic
University, Pomona CA 91768}

\newpage

\section{Introduction}
The nucleon resonance $S_{11}(1535)$ has the outstanding property of a very
strong $\eta N$ decay channel. The proper understanding of the low  energy
eta-nucleon interaction and threshold eta production consequently  hinges on a
correct description of the $S_{11}(1535)$ resonance. Recently precise eta
photoproduction data off protons and nuclei close to threshold have become
available. At MAMI (Mainz) \cite{krusche} very precise angular distributions
for the reaction $\gamma p \to \eta p$ have been measured from threshold at
$707$ MeV up to $800$ MeV photon lab energy. Together with an analogous
experiment using virtual photons (electroproduction) performed at ELSA (Bonn)
\cite{schoch} the data cover the whole range of the $S_{11}(1535)$ resonance.
The measured cross sections clearly exhibit the strong $S_{11}(1535)$
dominance of near threshold $\eta$-production. Furthermore, the cross sections
for the $\eta$-photoproduction off nuclei \cite{stroeher} show an
$A^{2/3}$-dependence on the mass number $A$, characteristic of strong,
surface-dominated interactions. Certainly, these new data demand
a closer look at this particular nucleon resonance.

The traditional picture of the $S_{11}(1535)$ is that of an excited three
quark  nucleon
resonance, with one of three quarks orbiting in an $l=1$ state around the
other two. This approach has however difficulties in explaining the large
$(30 - 55 \%$) $\eta N$ branching ratio. In \cite{isgur} the tensor force
from the hyperfine interactions due to one-gluon exchange can produce the
required SU(6) mixing to cause a large coupling to the $\eta N$ channel for
the $S_{11}(1535)$ and a near-zero coupling for the $S_{11}(1650)$. However,
then there are problems in reproducing the observed total decay width.

At present a frequently used ansatz for incorporating the $S_{11}(1535)$
resonance into the $\pi N$, $\eta N$ (and $\gamma N$) system is to couple these
channels directly to the $S_{11}(1535)$ via a phenomenological isobar model
\cite{liu}, \cite{bennh} with background terms \cite{benmer,sauermann}. In
these models, the coupling
constants $g_{\pi N N^*}$ and $g_{\eta N N^*}$ are treated as free parameters.
Their values vary in the literature, but always $g_{\eta N N^*}$ is the
larger of the two. The physical reason behind the large coupling of the
$S_{11}(1535)$ to the  $\eta N$ channel is not understood.

In this letter, we investigate the possibility that the $S_{11}(1535)$ is a
quasi-bound  meson-baryon S-wave resonance. The basis of our calculation is the
$SU(3)$ effective chiral lagrangian, with explicit symmetry breaking due to the
non-vanishing up, down and strange quark masses properly incorporated. This
approach successfully describes the
$\Lambda(1405)$ resonance as a quasi-bound $\overline{K} N$ state
\cite{kaiser}. Using the same lagrangian parameters as determined from our
$\overline{K} N$ analysis, we extend the formalism to the energy range $1480$
MeV $ < \sqrt{s} < 1600$ MeV to explore whether any $l=0$, $I=1/2$ resonances
can be  formed, and what their properties are. Indeed, we find a resonant state
with a large $\eta N$ decay width as well as other characteristic
properties of the $S_{11}(1535)$.

\section{Effective Chiral Lagrangian and Pseudo-Potential Approach}

The effective chiral lagrangian for meson-baryon interaction can be
systematically expanded in powers of small external momenta \cite{bkmrev}
\begin{equation}
  {\cal L} = {\cal L}^{(1)} + {\cal L}^{(2)} + \cdots
\end{equation}
where the superscript denotes the power of the meson momentum appearing in each
term. In the heavy  baryon mass formalism \cite{jenk}, the leading piece
at order $q$ reads
\begin{equation}
  {\cal L}^{(1)} = {\rm tr}( \overline{B}i v \cdot D B )
\end{equation}
with the chiral covariant derivative $D^\mu B = \partial^\mu B +
[\Gamma^\mu,B]$. As this part of the lagrangian incorporates all
current-algebra results of the meson-baryon interaction, it is refered to as
the Weinberg-Tomozawa or current algebra term. At next order in the expansion
scheme, $q^2$, there is a host of new terms allowed by chiral symmetry
\cite{krause}. In the heavy baryon formalism the most general form
relevant to S-wave scattering is given by \cite{kaiser}
\begin{eqnarray}
{\cal L}^{(2)} &=& {1 \over {2M_0}} {\rm tr}(\overline{B}((v \cdot D)^2 -
D^2) B)                      \nonumber  \\
               & & + b_D {\rm tr}(\overline{B} \lbrace \chi_+ , B \rbrace)+b_F
{\rm tr}(\overline{B} \lbrack \chi_+ , B \rbrack)  + b_0{\rm t r}(\overline{B}
B) {\rm tr}(\chi_+) \nonumber \\
    & & + d_D {\rm tr}(\overline{B} \lbrace (A^2 + (v \cdot A)^2),B \rbrace) +
d_F {\rm tr}(\overline{B} \lbrack (A^2 + (v \cdot A)^2),B \rbrack) \nonumber \\
        & & +d_0 {\rm tr}(\overline{B} B) Tr(A^2 + (v \cdot A)^2)   \\
    & & + d_1 ({\rm tr}(\overline{B} A_\mu) {\rm tr} (A^\mu B) + {\rm tr}
(\overline{B} (v \cdot A)) {\rm tr}((v \cdot A) B) ) \nonumber \\
    & &  + d_2 {\rm tr}(\overline{B} (A_\mu B A^\mu + (v \cdot A) B (v \cdot
    A))) \,.            \nonumber
\end{eqnarray}
The first term above is a relativistic correction involving the baryon mass
$M_0$ in the chiral limit. The parameters $b_D= 0.066$ GeV$^{-1}$ and $b_F=
-0.213$ GeV$^{-1}$  are determined from the mass splittings in the baryon
octet.   The other six parameters have been determined in a fit to the low
energy $\overline{K} N$ experimental data \cite{kaiser}, constrained by some
$\pi N $ and $KN$ data.

In order to investigate the possibility of resonance formation, one needs a
non-perturbative approach which resums a set of diagrams to all orders. Since
this leads necessarily beyond the systematic expansion scheme of chiral
perturbation theory, we use a potential model. A pseudo-potential is
constructed such that in the Born approximation it has the same
S-wave scattering length as the effective chiral lagrangian, at order $q^2$.
We note that this approach is quite similar to the one used in \cite{vkolck}
for the nucleon-nucleon interaction.

As in \cite{kaiser} we examine two ways of parameterizing the finite range
of the potential while keeping the Born term the same: a local potential and
one separable in the incoming and outgoing center-of-mass momenta. The local
potential between channels i and j is chosen to have a Yukawa form
\begin{equation}
V_{ij}(\vec{r}) = {{C_{ij} \alpha^2_{ij}} \over {8 \pi f^2}} \sqrt{{{M_i M_j}
  \over  {s \omega_i \omega_j }}}  {{e^{-\alpha_{ij} r}} \over {r}}
\end{equation}
where the indices $i,j \in \{1,2,3,4\}$ label the four channels $\pi N$, $\eta
N$, $K\Lambda$ and $K\Sigma$ respectively. $M_i$ and $\omega_i$ stand for the
baryon mass and reduced meson-baryon energy in channel $i$, $s$ is the squared
center-of-mass total energy and $f= 92.4 \pm 0.3$ MeV the pion decay constant
\cite{pdgroup}.
The  parameters  $\alpha_{ij}$ can be interpreted as average "effective masses"
representing the spectrum of exchanged particles in the $t$-channel mediating
the interaction. The relative interaction strengths $C_{ij}$ which follow
directly from the effective chiral lagrangian are listed in the appendix.

The potential of Eq.(4) is then inserted into the coupled channel
Schr\"odinger equation
\begin{equation}
(\nabla^2 + k_i^2 )\psi_i = 2 \omega_i \sum_{j=1}^4 V_{ij } \psi_j
\end{equation}
to solve for the multi-channel $S$-matrix. For comparison we also examine a
separable potential in momentum space, $\tilde V_{ij}(k_i,k_j)=
C_{ij}\sqrt{M_iM_j/s \omega_i \omega_i} \, \alpha_i^2 \alpha_j^2  [4 \pi^2 f^2
(\alpha_i^2 + k_i^2)(\alpha_j^2 + k_j^2)]^{-1}$, which is iterated in a
corresponding  Lippmann-Schwinger  equation as  described in \cite{kaiser}.

\section{Discussion and results}

It is instructive first to discuss qualitative aspects of the calculation.
The situation in the strangeness $S=0$ sector at energies near the $K \Sigma$
threshold is similar to the $S=-1$ case near the $\overline{K} N$ threshold.
There the  $\Lambda (1405)$ resonance can be produced as a quasi-bound
$\overline{K} N$  state resulting from the strong $I=0$ attraction between the
anti-kaon and the nucleon, as well as between the pion and sigma hyperon. This
attractive interaction  comes at leading order $q$ from the current algebra
term, Eq.(2). For the quantum numbers $S=0$, $I=1/2$ and $l=0$ there are
several important features:
\begin{itemize}
\item{} There is a strong attraction between the kaon and sigma (see the large
negative coefficient $C_{44}$ in the appendix). Thus, as soon as the inverse
range parameter $\alpha$ exceeds a certain minimal value, a bound state will be
necessarily formed below the $K \Sigma$ threshold.
\item{} The direct interaction between the $\eta$ meson and the nucleon is
very weak and there is a small direct coupling between the $\pi N$ and $\eta
N$  channels ($C_{22}$ and $C_{12}$ are small).
\item{} However, there is a strong coupling of both the $\pi N$ and $\eta N$
channels to the $K \Sigma$ channel ($C_{14}$ and $C_{24}$ are sizable).
\item{} Thus {\it the resonance formed will strongly connect the $\pi N$ and
$\eta N$ channels through the coupled channel dynamics}.
\end{itemize}

Let us for the moment consider only the current algebra piece, $i.e.$ all $b$-
and $d$-parameters are zero and the $1/M_0$ corrections are neglected. In
this case $C_{22} = C_{12} = 0$, but $C_{23}$ and $C_{24}$ are large.
In Table 1 we show the resonance energy versus $\alpha$ for both the local and
separable potential using a common inverse range $\alpha$ for all channels. The
resonance position is identified when an eigenphase of the multi-channel
$S$-matrix is equal to 90 degrees.  Thus, if $\alpha > 490$ MeV for the local
potential (or $670$ MeV for the separable one) a resonance is necessarily
formed below the $K \Sigma$ threshold from the current algebra piece alone.
Experimentally, there are two $S_{11}$ nucleon resonances in this energy range,
at $1535$ and $1650$ MeV. Since only the $S_{11}(1535)$ has a large $\eta N$
branching ratio it is the main candidate for this dynamically generated
resonance.

Next we include all order $q^2$ terms using values of the $b$- and
$d$-parameters as previously obtained from a fit to the low energy
$\overline{K} N$ data and allow for a $\pm 5\% $ uncertainty in the
parameters. We note that  they are  similar for both  potential  forms
\cite{kaiser}. Thus the only free parameters are the $\alpha_{ij}$ in
Eq.(4). Since the $\pi N$ channel is far above its threshold a satisfactory
fit to all the data using only one common range for all channels could not
possibly be expected. However, a good fit was found using
only two range parameters: one for the $\pi N$ channel, and one common range
for the other three. The off-diagonal ranges were taken to be $\alpha_{ij} =
(\alpha_i + \alpha_j)/2$.

We performed a coupled  channel calculation for the $\pi N$ $S_{11}$ phase
shift and inelasticity, as well as the $\pi^- p \to \eta n$
cross section. The results of the fit for both the local and separable
potential forms are shown in Figs.1 and 2a,b. Here the range parameters are
$\alpha_{\pi N} = 320$ MeV and $\alpha_{\eta N}=\alpha_{K \Lambda}=\alpha_{K
\Sigma}=530$ MeV for the local potential.  For the separable potential the
range parameters are $\alpha_{\pi N}=573$ MeV and $\alpha_{\eta N}=\alpha_{K
\Lambda}=\alpha_{K \Sigma}=776$ MeV. The values of $b_0$ and the $d$
parameters, which only differ by $5 \%$ from \cite{kaiser}, are listed in
Table 2.

It is remarkable that such a good fit to the $\pi N$ $S_{11}$ phase shift and
the $\eta$-production cross section is obtained with only two free parameters.
Clearly, one can not expect the $S_{11}$ inelasticity to be accurate since the
$\pi\pi N$ channel is neglected here. Nevertheless, this picture of the
$S_{11}(1535)$ as a dynamic resonance based on the effective chiral lagrangian
reproduces many of its properties. For example we obtain a resonance mass $M^*
= 1557$ MeV and a full width $\Gamma_{tot} = 179$ MeV. These values agree
favorably with existing empirical determinations \cite{pdgroup},\cite{krusche}.
As byproduct we extract the $\eta N$ S-wave scattering length to be $a_{\eta N}
= (0.68 + i \, 0.24)$ fm. This number is close to values found from other
analyses \cite{wilkin,sauermann,arima}.

In Fig.3 we display the $K\Sigma$ and $K\Lambda$ components of the bound state
wave function at resonance. The root mean square radii are $0.70$ fm and $0.88$
fm for the $K\Sigma$ and $K\Lambda$ components.

\section{Resonance and background effects}

If there are only two reaction channels, it is often useful to parameterize the
$S$-matrix  in terms of its two eigenphases and a mixing angle $\epsilon$.
In the case of a pure Breit-Wigner resonance  the $T$-matrix has the following
energy dependence (on $W=\sqrt{s}$):
\begin{equation} T(W) = {1 \over 2(M^*- W) - i\,\Gamma(W) }
  \left( \begin{array}{cc} \gamma_1 & \sqrt{ \gamma_1 \gamma_2} \\
  \sqrt{\gamma_1 \gamma_2} &   \gamma_2 \end{array} \right)\,,
\end{equation}
with det$T(W) =0$. The constant $M^*$ is the resonance mass and $\Gamma(W)$
the (energy dependent) width. For a S-wave resonance which decays into
two-particle final states unitarity requires the energy dependence of the width
to be  $\Gamma(W) = \gamma_1 \,k_1(W) + \gamma_2\, k_2(W)$. Here, $k_i(W)$ is
the center-of-mass momentum in channel $i$ and the constants $\gamma_i$  are
related to  the partial decay widths $\gamma_i \, k_i(M^*)$.
For a pure Breit-Wigner resonance, one eigenphase of the $S$-matrix
(background) is zero and the other one (resonant) has the energy dependence
\begin{equation}
  \tan\delta_{res}(W) = {\Gamma(W)\over 2(M^*-W)}\,.
\end{equation}
Even though we do not have a pure Breit-Wigner resonance we find a resonant
eigenphase (see Fig.4) which is very close to a Breit-Wigner form. In this
figure, we plot both the resonant and non-resonant eigenphases versus pion lab
kinetic energy for our calculation. The dots correspond to a Breit-Wigner form
with parameters $M^*=1557$
MeV, $\gamma_\pi = 0.26$ and $\gamma_\eta =0.25$. These numbers result in
partial decay widths $\Gamma_\pi = 124$ MeV and $\Gamma_\eta = 55$ MeV. The
branching ratio $b_\eta = 0.31 $ is still compatible with the existing analysis
\cite{pdgroup} whereas $b_\pi = 0.69$ is somewhat too large, presumably due to
the neglect of the $\pi\pi N$ channel and our way of extracting $\gamma_i$.
Here, the $\gamma_i$ are determined from the energy dependence of
the resonant eigenphase. We note furthermore that at the $\eta
N $-threshold ($W_{th} = M_N + m_\eta$) the $\pi N$ $S_{11}$ phase shift reads
according to Eq.(7)
\begin{equation}
\tan \delta_{11}(W_{th}) = {\gamma_\pi k_\pi(W_{th}) \over 2
(M^*-W_{th}) }  \end{equation}
within the two-channel calculation, since the background phase is zero at
$W_{th}$. Therefore a good knowledge of this
particular phase constrains the resonance mass and the $\pi N$ partial decay
width.

 From the $T$-matrix for a pure Breit-Wigner resonance in Eq.(6), the
ratio of cross sections for scattering from channel 1 $\to$ 2 divided by that
for elastic scattering (1 $\to$ 1) is $\sigma_{21}(W)/\sigma_{11}(W) =
\gamma_2\,k_2(W)/ \gamma_1\, k_1(W)$. Therefore, the ratio $R_{BW}$ defined as
\begin{equation}
   R_{BW} \equiv { \gamma_1\,k_1(W) \,\sigma_{21}(W) \over  \gamma_2
\,   k_2(W)\,\sigma_{11}(W)} \,,\end{equation}
is exactly one for a pure Breit-Wigner resonance. Any deviation from unity
originates from the background eigenphase, assuming that the resonant
eigenphase has well determined partial widths.  In Fig. 5 we plot the quantity
\begin{equation}
  R_{BW} = {{\gamma_{\pi} k_\pi \,\sigma(\pi N \to \eta N,S_{11})} \over
        {\gamma_{\eta } k_\eta \,\sigma(\pi N \to \pi N,S_{11})}}
\end{equation}
which involves $S_{11}$ partial wave cross sections only, versus the pion lab
kinetic energy. The solid line corresponds to the potential model used
here. Since the $\eta$-production
near threshold is strongly S-wave dominated, one can identify $\sigma(\pi N
\to \eta N,S_{11})$ with ${3\over 2} \sigma(\pi^-p \to \eta n)$. The $S_{11}$
component of the elastic $\pi N$ cross section can be constructed from the
partial wave analysis of \cite{hohler,arndt}. Using these inputs together with
$\gamma_\pi/\gamma_\eta= 1.04$ as determined from the shape of our resonant
eigenphase, we display the result for this ratio. Since presently the branching
ratios $b_\pi, b_\eta$ have large uncertainties, we choose for reasons of
comparison the values obtained here. The error bars in Fig.5 reflect only those
of $\eta$-production data. It is visible that $R_{BW}$ deviates from unity
as required for a pure Breit-Wigner resonance. This is an indication that
background effects (corresponding to a non-resonant eigenphase) are not
negligible. In \cite{sauermann} a similar phenomenon was observed in sofar as
the coupling constants of the resonance  depended strongly on the treatment of
the background (nonresonant) amplitude. Also in \cite{oset} the coupling
constants of the $S_{11}(1535)$ had to be varied  up to $20 \%$ from the values
obtained from the widths in order to obtain a  good fit the the scattering
data. This all points towards the presence  of some background amplitude.
We remark however that inclusion of the $\pi\pi N$ channel in our analysis
would change the branching ratios and could considerably lower the ratio
$R_{BW}$ from the value shown in Fig.5. If possible an experimental
determination of this ratio would be very valuable.

In summary, we have used the effective chiral lagrangian at next-to-leading
order to investigate the possibility that the $S_{11}(1535)$ resonance is a
quasi-bound $K \Sigma$-$K \Lambda$ state. Using the same parameters as
obtained from fitting the low energy $\overline{K} N$ data and two free finite
range parameters, a resonance can be formed at $1557$ MeV with the
characteristic properties of the $S_{11}(1535)$. The $\pi N$ $S_{11}$ phase
shifts and inelasticities as well as the $\eta N$-production cross section are
remarkably well reproduced. The dynamically generated resonance has a full
width of $179$ MeV and branching ratios extracted from the shape of the
resonance of $69\%$ into $\pi N$ and $31\%$ into $\eta N$ final
states. Furthermore, we elaborated on the background  effects in the reactions
dominated by the $S_{11}(1535)$ and proposed as a measure for it the ratio
$R_{BW}$ in Eq.(10). The coupled  channel approach
presented here can also be used in calculations of the $\eta$-photoproduction
process, and  we hope to report on this topic in the near future.

\bigskip

\centerline{\bf Appendix }

\bigskip

Here we list the expressions of the relative coupling strengths $C_{ij}$ in the
$I=1/2$ basis entering the potential of Eq.(4) in terms of the chiral
lagrangian parameters. The indices $1,2,3,4$ refer to the $\pi N$, $\eta N$, $K
\Lambda$, $K \Sigma$ channel respectively and the $\eta$-particle is
identified with the $SU(3)$-octet state $\eta_8$.  Furthermore, $E$ denotes the
center-of-mass meson energy and $M_0 \simeq 0.91$ GeV is the octet baryon mass
in the chiral limit.
\begin{eqnarray}
   C_{11} & = & -E_\pi  + {1 \over {2 M_0}}(m_\pi^2 - E_\pi^2)
     + 2 m_\pi^2(b_D + b_F + 2 b_0) - E_\pi^2 (d_D + d_F + 2 d_0) \nonumber \\
   C_{12} & = &  2 m_\pi^2 (b_D + b_F) + E_\pi E_\eta
                  (d_2 - d_D - d_F) \nonumber \\
   C_{13} & = &  {3 \over 8} (E_\pi + E_K) + {3 \over {16M_0}}
                 (E_\pi^2-m_\pi^2+E_K^2-m_K^2) - {1 \over 2}(m_K^2+m_\pi^2)
                 (b_D+3b_F)  \nonumber \\
          &   &   + {{E_\pi E_K} \over 2} (d_D+3d_F-d_2) \nonumber \\
   C_{14} & = &   -{1 \over 8}(E_\pi+E_K) - {1 \over {16 M_0}}
                  (E_\pi^2-m_\pi^2+E_K^2-m_K^2) + {1 \over 2}(b_F -b_D)
                  (m_\pi^2+m_K^2)  \nonumber \\
          &   &  + {{E_\pi E_K} \over 2} (d_D-d_F-2d_1-3d_2)   \nonumber \\
   C_{22} & = &  {16 \over 3} m_K^2(b_D-b_F+b_0) + 2 m_\pi^2({5 \over 3} b_F
                         - b_D-{2 \over 3}b_0)+E_\eta^2
                   (d_F-{5 \over 3}d_D-2d_0+{2 \over 3} d_2) \\
   C_{23} & = &  {3 \over 8}(E_\eta +E_K ) + {3 \over {16M_0}}
                 (E_K^2-m_K^2+E_\eta^2-m_\eta^2) + (b_D+3 b_F)({5 \over 6}
                m_K^2-  {1 \over 2}m_\pi^2) \nonumber \\
          &   &  -E_\eta E_K({d_F \over 2} +
                  {d_D \over 6} + d_1 + {{5d_2} \over 6} )\nonumber \\
   C_{24} & = &   {3 \over 8} (E_\eta + E_K) + {3 \over {16M_0}}
                 (E_K^2-m_K^2+E_\eta^2-m_\eta^2) + ({5 \over 2}m_K^2-
                   {3 \over 2}m_\pi^2)(b_F-b_D) \nonumber  \\
          &   &   + {{E_\eta E_K} \over 2}
                    (d_D-d_F-d_2) \nonumber \\
   C_{33} & = &   ({10 \over 3}b_D+4b_0) m_K^2+ E_K^2
                 ({{2d_2} \over 3}-2d_0-{{5d_D} \over 3}) \nonumber\\
   C_{34} & = &   2m_K^2b_D+E_K^2(d_2-d_D) \nonumber \\
   C_{44} & = &   -E_K -{1 \over {2 M_0}}(E_K^2-m_K^2) +
                  2m_K^2(b_D-2b_F+2b_0)  + E_K^2(2d_F-d_D-2d_0) \nonumber
\end{eqnarray}

\bigskip

\begin{center}
\begin{tabular}{|cc|cc|} \hline
\multicolumn{2}{|c|}{Local Potential} & \multicolumn{2}{c|}{Separable
Potential} \\
$\alpha$ (MeV) & Energy & $\alpha$ (MeV) & Energy \\
\hline
490 & 1661 & 670 & 1661 \\
520 & 1604 & 710 & 1604 \\
550 & 1550 & 750 & 1556 \\
575 & 1489 & 760 & 1501 \\
\hline
\end{tabular}
\end{center}
\vspace{.2in}
\noindent {\bf Table 1}.  The energy of the $K \Sigma$-$K \Lambda$  ($I=1/2$)
quasi-bound state produced from the current algebra (Weinberg-Tomozawa) term
alone as a function of the range parameter $\alpha$ for both the local and the
separable potential.  The range parameter $\alpha$ is the same for all
channels.

\bigskip

\begin{center}
\begin{tabular}{|c|cccccc|cc|}\hline
Potential & $b_0$ & $d_0$ & $d_D$ & $d_F$ & $d_1$ & $d_2$ & $\alpha_{\pi N}$ &
 $\alpha_{K \Sigma}$  \\
\hline
Local   & --0.517 & --0.68 & --0.02 & --0.28 & +0.22 & --0.41 & 0.32  & 0.53 \\
Separable & --0.279 & --0.42 & --0.23 & --0.41 & +0.27 & --0.65 & 0.57 & 0.77
\\ \hline \end{tabular}
\end{center}
\noindent{\bf Table 2}. Values of the Lagrange parameter entering at order
$q^2$ in units of GeV$^{-1}$. The inverse ranges $\alpha$ are given in GeV.

\bigskip

\begin{center}
Figure Captions
\end{center}
\medskip
\begin{itemize}
\item[Fig.1] The cross section $\sigma(\pi^- p \to \eta n)$ versus the
pion lab kinetic energy $T_\pi$. The selected data are taken from
\cite{nefkens}. The solid/dashed line corresponds to the local/separable
potential form.
\medskip

\item[Fig.2a] The pion-nucleon $S_{11}$ phase shift as a function of the pion
lab  kinetic energy. The triangles/circles are from the phase shift analysis
of \cite{hohler}/ \cite{arndt}. The full/dashed curve corresponds to a
calculation using a local/separable potential form.
\medskip

\item[Fig.2b] The pion-nucleon $S_{11}$ inelasticity as a function of the pion
lab kinetic energy. The notation is the same as in Fig.2a.
\medskip

\item[Fig.3] The two-component bound state wave function at resonance versus
the meson baryon distance $r$.
\medskip

\item[Fig.4] The eigenphases of the multi-channel $S$-matrix below the
$K\Lambda$-threshold. The heavy dots correspond to a Breit-Wigner fit of the
resonant phase.
\medskip

\item[Fig.5] The ratio $R_{WB}$ defined in Eq.(10). The notation is the same as
in Fig.2a. The error bars reflect only those of the $\eta$-production cross
sections.
\medskip

\end{itemize}

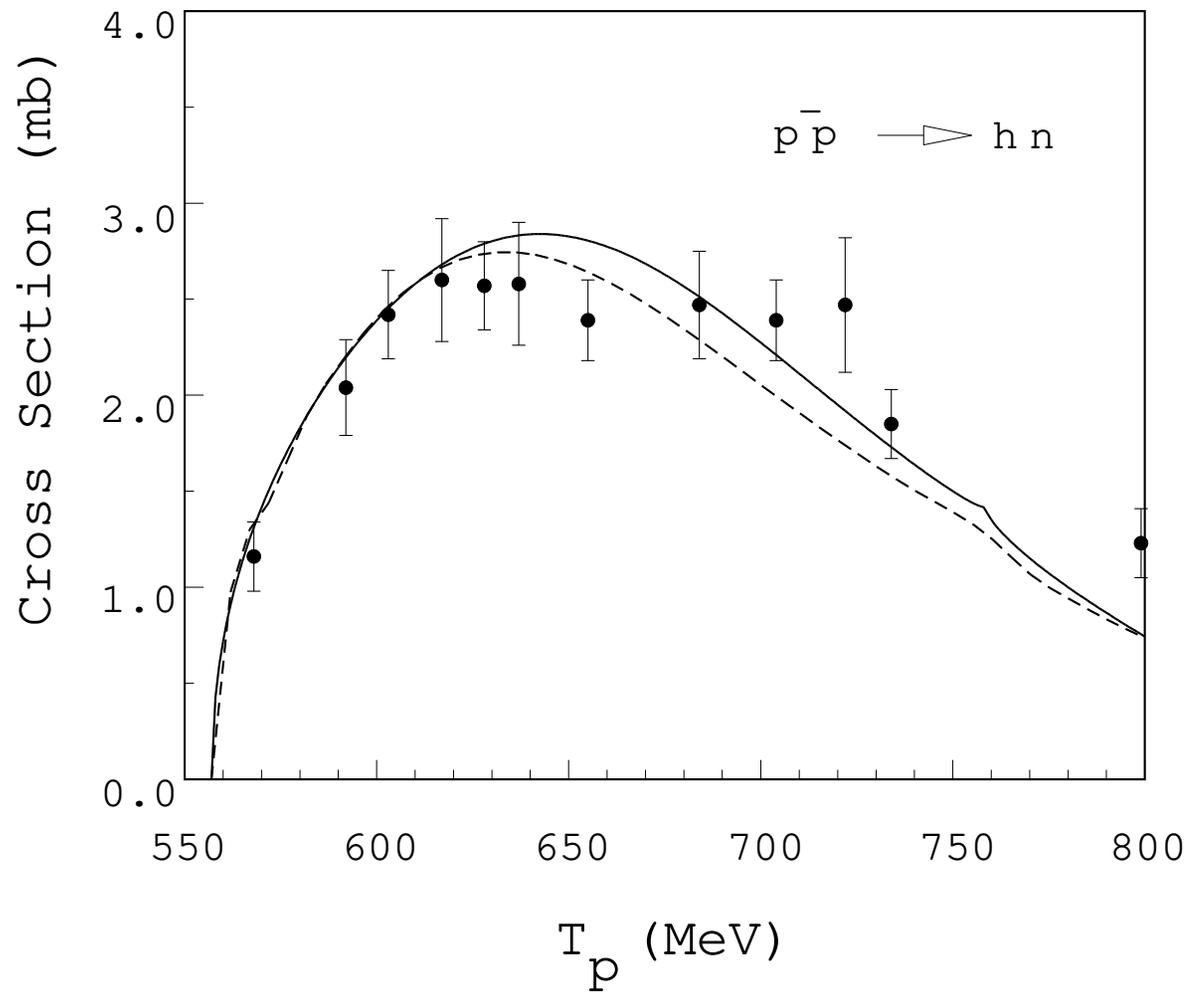
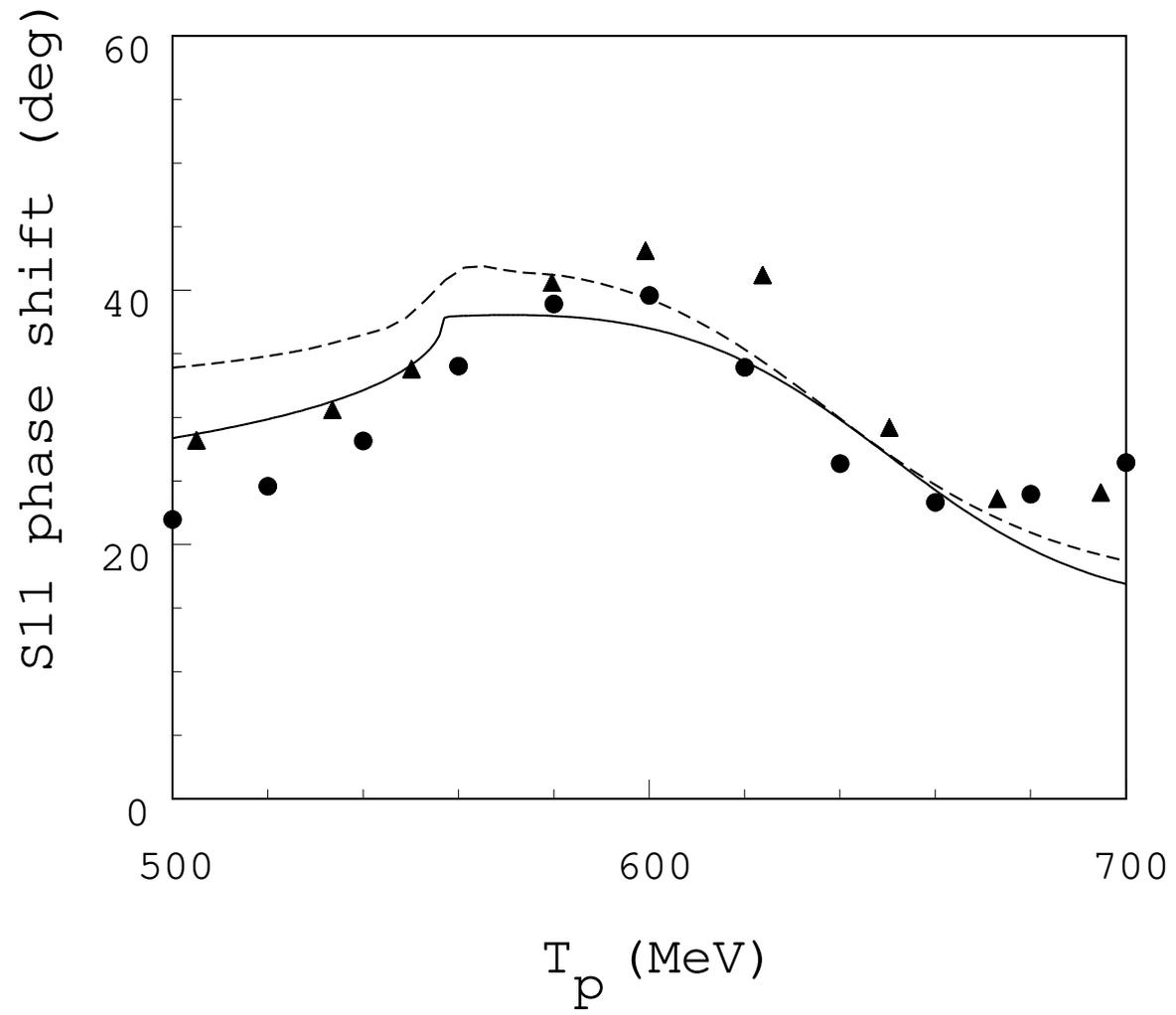
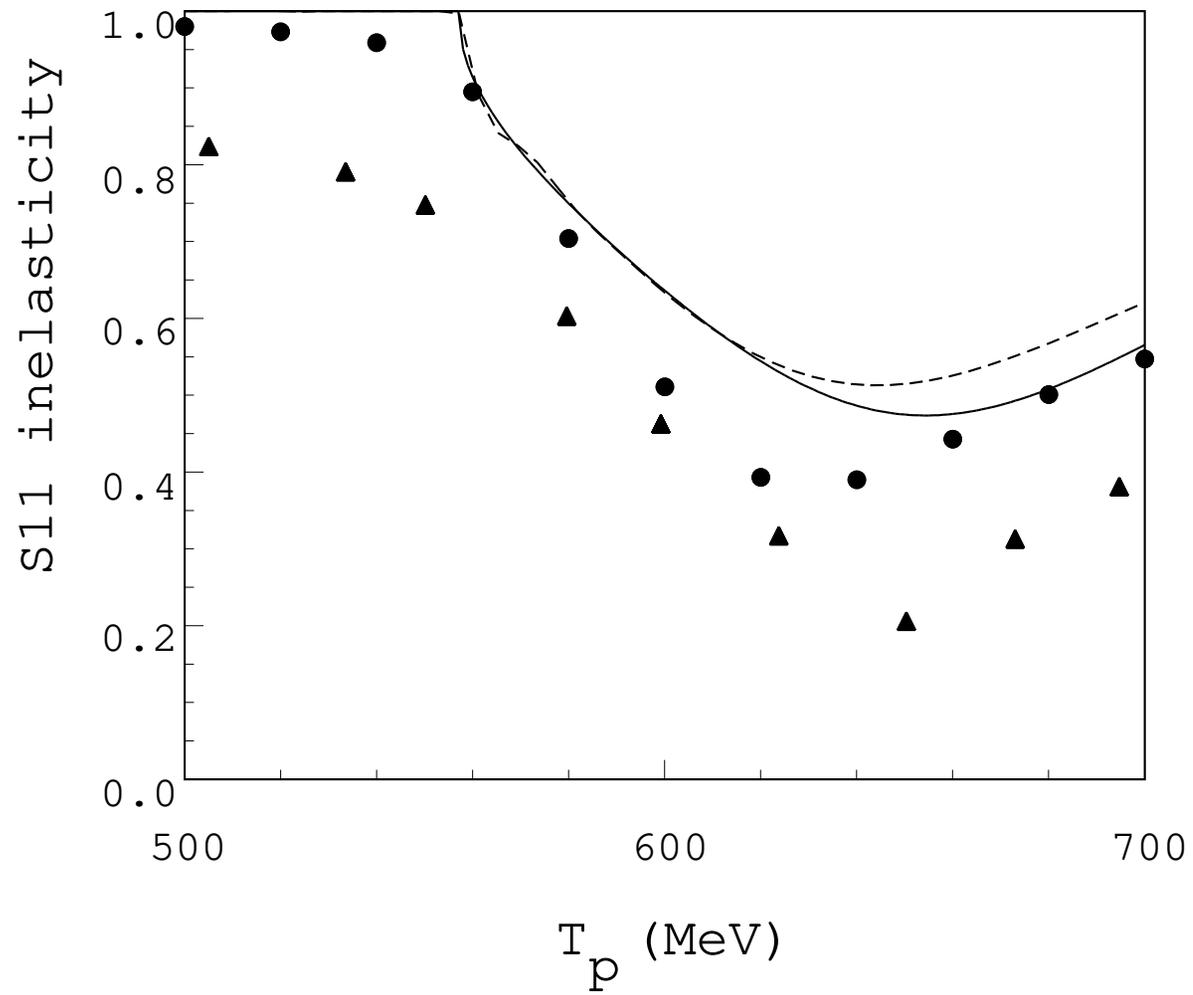
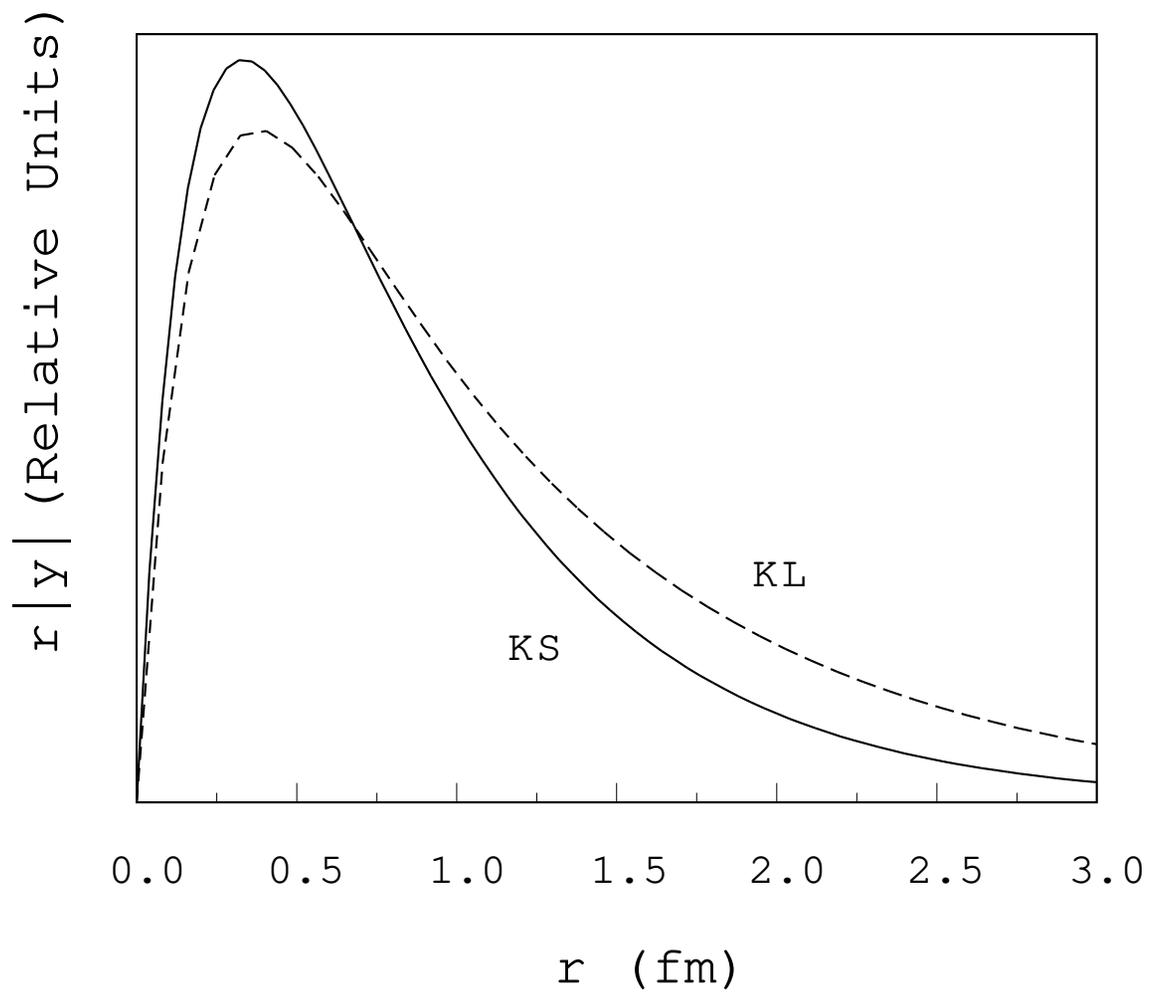
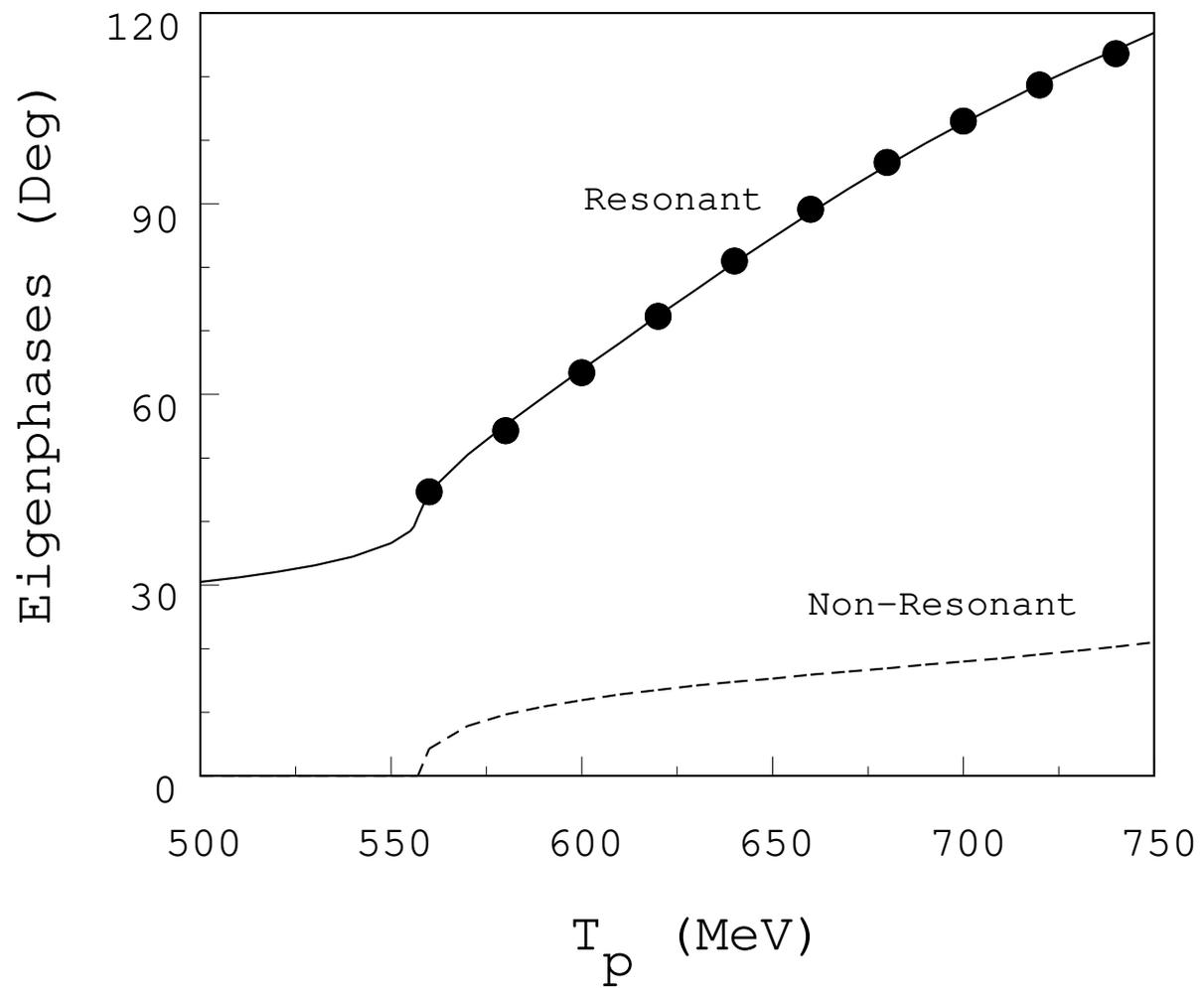
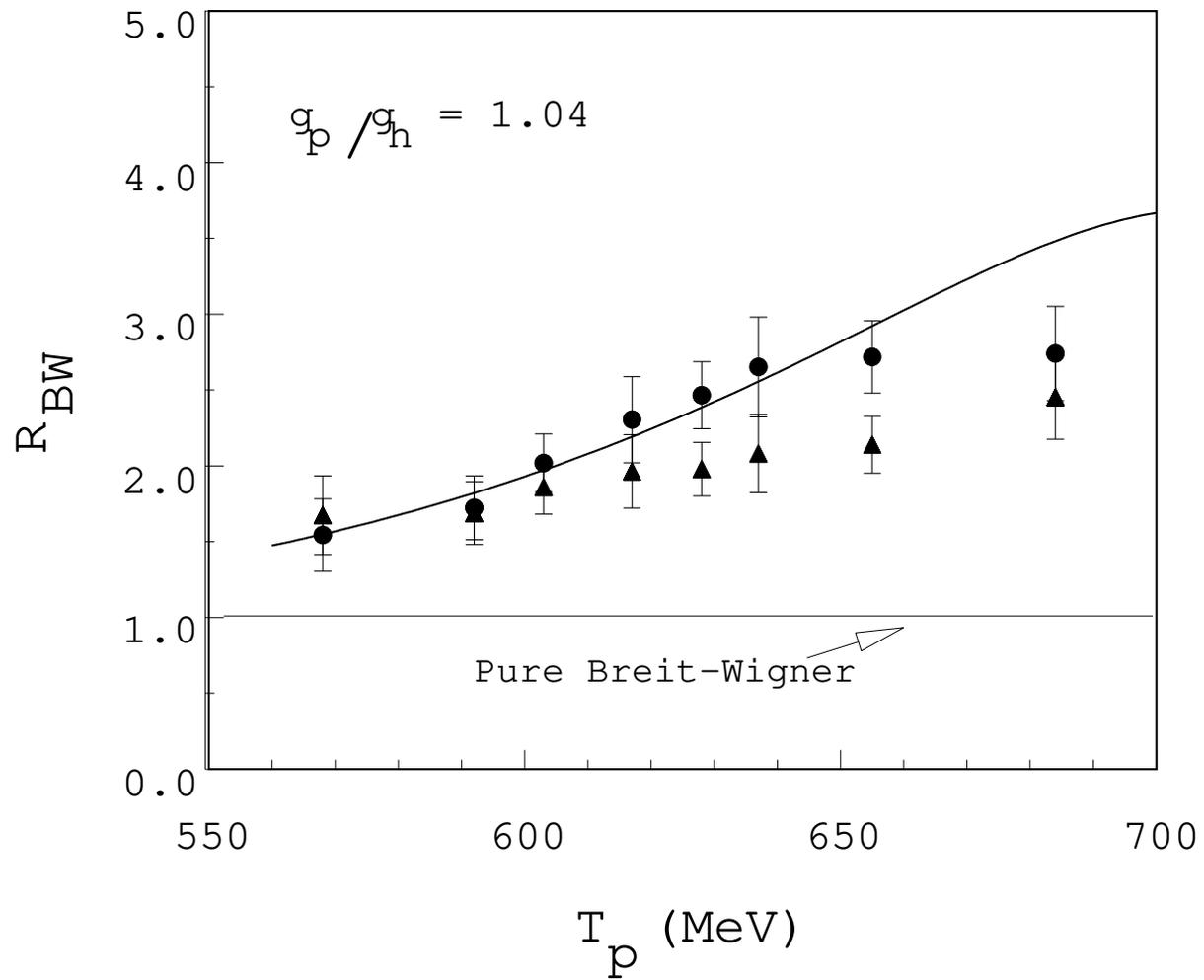
\end{document}